\documentclass[journal]{IEEEtran}
\usepackage{booktabs}
\usepackage{xcolor,soul,framed}
\usepackage{float} 
\usepackage[margin=1in]{geometry}
\colorlet{shadecolor}{yellow}
\usepackage{graphicx}
\graphicspath{{../pdf/}{../jpeg/}}
\DeclareGraphicsExtensions{.pdf,.jpeg,.png}
\usepackage{caption}
\usepackage[export]{adjustbox}
\usepackage{wrapfig}
\usepackage[skip=2pt,font=footnotesize]{caption}
\usepackage{hyperref}

\hypersetup{
    colorlinks=true,
    linkcolor=blue,
    filecolor=blue,      
    urlcolor=black,
    citecolor =blue,
    pdftitle={Permittivity Characterization of 3D Printed Materials in Millimeter Waves},
    pdfpagemode=FullScreen
    }
\usepackage{subfig}
\usepackage[cmex10]{amsmath}
\usepackage{array}
\usepackage{mdwmath}
\usepackage{mdwtab}
\usepackage{eqparbox}
\usepackage{url}
\usepackage{textcomp}
\usepackage{gensymb}
\usepackage{multirow}
\usepackage{multicol}
\usepackage{stfloats}
\usepackage{mathrsfs}
\usepackage[noadjust]{cite}
\usepackage[ragged]{sidecap}
\usepackage{bm}

\begin{document}
\title{Permittivity Characterization of 3D-Printed Materials at Millimeter Waves}

\author{
Kamil A. Işık{*}$^{1,2}$, Mohammad M. Asgari{*}$^1$,
Xuchen Wang$^3$,\\
Maxim Masyukov$^1$,
Irina Nefedova$^1$,
Zachary Taylor$^1$,
and Viktar~S.~Asadchy$^1$ \\

    $^1$Aalto University, Department of Electronics and Nanoengineering, Espoo 02150, Finland \\
    $^2$Department of Electrical and Electronics Engineering, METU, 06531, Ankara, Türkiye\\
    $^3$College of Physics and Optoelectronic Engineering, Harbin Engineering University, China
\thanks{{*}These authors contributed equally to this work.}

}  
\maketitle
\begin{abstract}
In this study, we characterize the permittivity of various commercially available materials commonly used in 3D printing within the millimeter-wave range of 70--110 GHz. The open-waveguide extraction method is employed to efficiently determine the permittivity of these 3D-printed materials, and we discuss key strategies to improve its accuracy. This methodology can be readily applied to the characterization of other similar materials and extended to higher frequency ranges. The resulting permittivity data are expected to support advancements in 6G and beyond wireless communication technologies, where 3D-printed materials and millimeter-wave frequencies are anticipated to play an increasingly significant role in the development of new antennas and metasurfaces.
\end{abstract}
\IEEEpeerreviewmaketitle
\section{Introduction}\label{sec:Intro}

\IEEEPARstart{K}nowledge of material permittivity is essential for numerous electromagnetic applications. Conventional materials generally have accurately measured permittivity over a wide frequency range \cite{johnson1972optical,palik1998handbook,aspnes1983dielectric,hagemann1975optical,werner2009optical}. Recently, advances in additive manufacturing have led to a new class of materials gaining significant attention, particularly dielectric materials used for 3D printing. These materials show great promise for electromagnetic applications, such as in antennas \cite{mirzaee2015developing,wang2017wideband}, lenses~\cite{zhang20163d,furlan20163d}, and metasurfaces \cite{jiang2019experimental,zhang2020vibration}.  The dielectric materials for 3D printing include a variety of polymers, such as acrylonitrile butadiene styrene (ABS), polylactic acid (PLA), nylon/polyamide, acrylonitrile styrene acrylate (ASA), polyethylene terephthalate (PET), polyethylene terephthalate glycol-modified (PETG), and polycarbonate (PC) \cite{wickramasinghe2020fdm} among others. Additionally, photopolymers, like standard resin, are widely utilized in resin-based 3D printing processes \cite{maines2021sustainable}. 

Recently, numerous studies have concentrated on measuring the permittivity of various 3D-printed dielectric materials. These studies can be broadly categorized into those employing resonant and transmission-line-based characterization methods. The resonant methods, such as cavity-based techniques, involve measuring the real and imaginary part of the permittivity from the resonance frequency and the Q-factor of a cavity, respectively. Such methods are well-suited for determining the permittivity of low-loss materials but are typically only suitable for measurements at discrete frequency points~\cite{769347,gershon2000adjustable,li2016measurement,cook1974comparison,1514643}. Transmission-line methods, on the other hand, are well-suited for broadband permittivity characterization. One common transmission-line approach is measuring materials' scattering matrix in free-space setups~\cite{seo2004characterization,4314617,aman2021free,jin2006terahertz,ruan2019terahertz,Duangrit2019,Busch2014,4014686}, which eliminates the need for physical contact with the material. However, free-space setups generally require sophisticated equipment and antennas that must be precisely calibrated for accurate results. Usually, these setups cover frequency ranges from 200~GHz and above~\cite{jin2006terahertz,ruan2019terahertz,Duangrit2019,Busch2014}. In addition, free-space measurements demand a large, clean, controllable measurement space to prevent reflections from surroundings. 
Therefore, another commonly used approach involves measuring the sample inside a closed waveguide. This method allows for accurate measurement of permittivity at the lower part of the spectrum, as the sample can be precisely shaped and positioned within the waveguide aperture. However, at higher frequencies, typically above 50~GHz, achieving a perfect fit within the waveguide becomes increasingly challenging due to the small waveguide dimensions~\cite{nicolson1970measurement,weir1974automatic,costa2017electromagnetic}.
While many studies have investigated the permittivity of various 3D-printed materials in the GHz and THz ranges using the above-mentioned methods \cite{Ruan2019,Duangrit2019,Jin2006,Dorozhkin2020,Busch2014,EscobariVargas2022,Wang2022,Huber2020,Boussatour2018,Sahin2019,Felicio2020,Meriakri2012,Cresson2014}, there appears to be a frequency gap between 50 GHz and 200 GHz wherein the measurement data were not retrieved. 
Nevertheless, this range is increasingly relevant for future wireless communication technologies, including 6G~\cite{di2020smart}. It is evident that in the near future a wide array of communication devices, such as antennas and metasurfaces, will rely on materials fabricated through cost-effective 3D printing technologies~\cite{asgari2024multifunctional}. 

In this paper, we perform permittivity extraction of various dielectric materials commonly exploited in 3D printing using the recently proposed open-waveguide method \cite{wang2022fast} within the frequency range of 70 GHz to 110 GHz. This method presents a rapid and robust approach for measuring dielectric slabs in a rectangular waveguide junction. It does not require precise control over the sample's shape or positioning, making it especially suitable for millimeter-wave and sub-terahertz measurements.  While the approach accurately determines the real part of permittivity, its estimation of the imaginary part is effective mainly for materials with medium to high losses. We present experimentally measured permittivity for several commonly used affordable materials, including seven different types of Ultimaker PLA and two resins produced by Formlabs. We also discuss how the utilized methodology can be modified to account for the unique properties of 3D materials. This methodology can be easily extended to other 3D materials and applied to higher frequency ranges, such as 110-200 GHz, with appropriate adjustments in sample thickness.


\section{Methodology}\label{sec:methodology}

\begin{figure}[tb]
\centering
\includegraphics[width=0.5\textwidth]{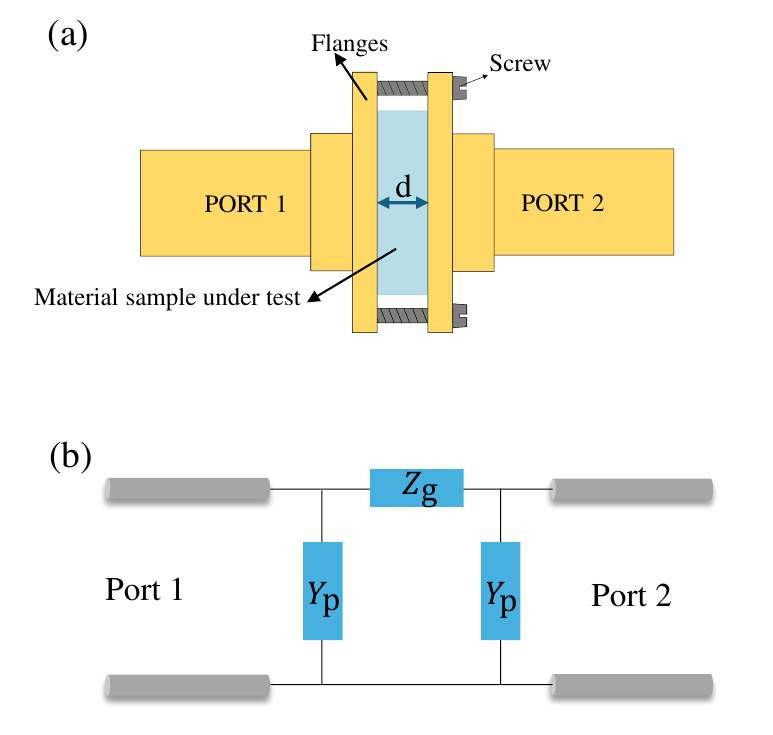}
\caption{(a) The schematic of the experimental setup illustrating the measurement configuration. (b) The equivalent circuit model of the experimental setup.}
\label{fig:setup}
\end{figure}

In this section, we present the method used to extract key parameters from the sample, incorporating both insights from prior work~\cite{wang2022fast}  and new findings from the current work. The schematic of the experimental setup can be seen in Fig.~\ref{fig:setup}(a). 
The sample of thickness $d$ is placed between two waveguide flanges, as shown in Fig.~\ref{fig:setup}(a), creating a discontinuity that is modeled as an equivalent $\pi$-circuit (Fig.~\ref{fig:setup}(b)). The waveguide flanges are tightened until they barely touch the sample, ensuring that the separation between the flanges is equal to $d$. 
The elements of the $\pi$-circuit can be extracted from the measured S-parameters. 
A key feature of this approach is that the equivalent shunt impedance of the sample is determined solely by its permittivity, making it insensitive to the sample's precise shape, size, or positioning. However, the method has limitations related to sample thickness, working best for thinner samples. For very thin slabs with thickness not exceeding $\lambda_{\rm d}/10$, where $\lambda_{\rm d}$ is the wavelength in the dielectric slab, an analytical formula can be directly applied to extract the permittivity from the measured $S$-parameters~\cite{wang2022fast}:
\begin{equation}
Y_{\rm p} = \frac{\cos(\beta_{\rm d} d) - 1}{j Z_{\rm d} \sin(\beta_{\rm d} d)} = \frac{1 - S_{11} - S_{21}}{Z_0 (1 + S_{11} + S_{21})},
\label{eq1}
\end{equation}
where   \( Y_{\rm p} \) is the shunt admittance shown in Fig.~\ref{fig:setup}(b),   \( \beta_{\rm d} = \sqrt{\omega^2 \mu_0 \varepsilon_0 \varepsilon_{\rm r} - \pi^2/L^2 }\) is the propagation constant in the material under test for the TE$_{10}$ mode, $\varepsilon_{\rm r}$ is the relative permittivity of the material, $L$ is the larger dimension of the waveguide aperture,
\( Z_{\rm d} =\mu_0 \omega/\beta_{\rm d} \) is the characteristic impedance of the waveguide for the TE$_{10}$   mode filled with the dielectric material under test, and     \( Z_0 \) is the characteristic impedance of the waveguide for the same mode. Indeed, by finding $\beta_{\rm d}$ from (\ref{eq1}), one can extract the unknown $\varepsilon_{\rm r}(\omega)$.

\begin{table}[H]
    \centering
    \caption{Thicknesses of measured material samples.}
    \begin{tabular}{@{}lc@{}}
        \toprule
        Material         & Average Thickness (mm) \\ \midrule
        Green PLA       & 0.694             \\
        Blue PLA        & 0.698             \\
        Red PLA         & 0.677             \\
        Silver PLA      & 0.693             \\
        White PLA       & 0.670             \\
        White Resin     & 0.754             \\
        Transparent PLA & 0.675             \\
         Black Resin    & 0.713             \\
        Black PLA       & 0.705             \\ \bottomrule
    \end{tabular}
    \label{tab:material_thicknesses}
\end{table}

For thicker material samples with thickness satisfying $\lambda_{\rm d}/10<d<\lambda_{\rm d}/2$, numerical optimization is required to fit the simulated data to the measured data~\cite{wang2022fast}. 
In this work, we operate within this range of sample thicknesses to ensure the sample is not fragile and to achieve better relative accuracy in measuring its thickness. For the operational frequency range of 70  -- 110 GHz, the maximum allowable thickness is approximately $d_{\max}=0.8$~mm. The thicknesses of the materials under test used in the measurements are listed in Table~\ref{tab:material_thicknesses}. 
While calculating average thicknesses of whole samples, standard deviation of the measured thicknesses are found approximately 0.015 mm.
We used WR10 waveguide with $2.54\,\text{mm} \times 1.27\,\text{mm}$
 aperture dimensions. 

The numerical optimization is based on the modeling of the measured sample with parametrized permittivity in numerical tools (ANSYS HFSS) and fitting the simulated equivalent shunt admittance $Y_{\rm p}$ with its measured values obtained using (\ref{eq1}). In this case, one does not need to accurately model the actual measurement setup which could introduce additional source of the error~\cite{nefedova2015dielectric}.  

\begin{figure}[tb]
\centering
\includegraphics[width=0.5\textwidth]{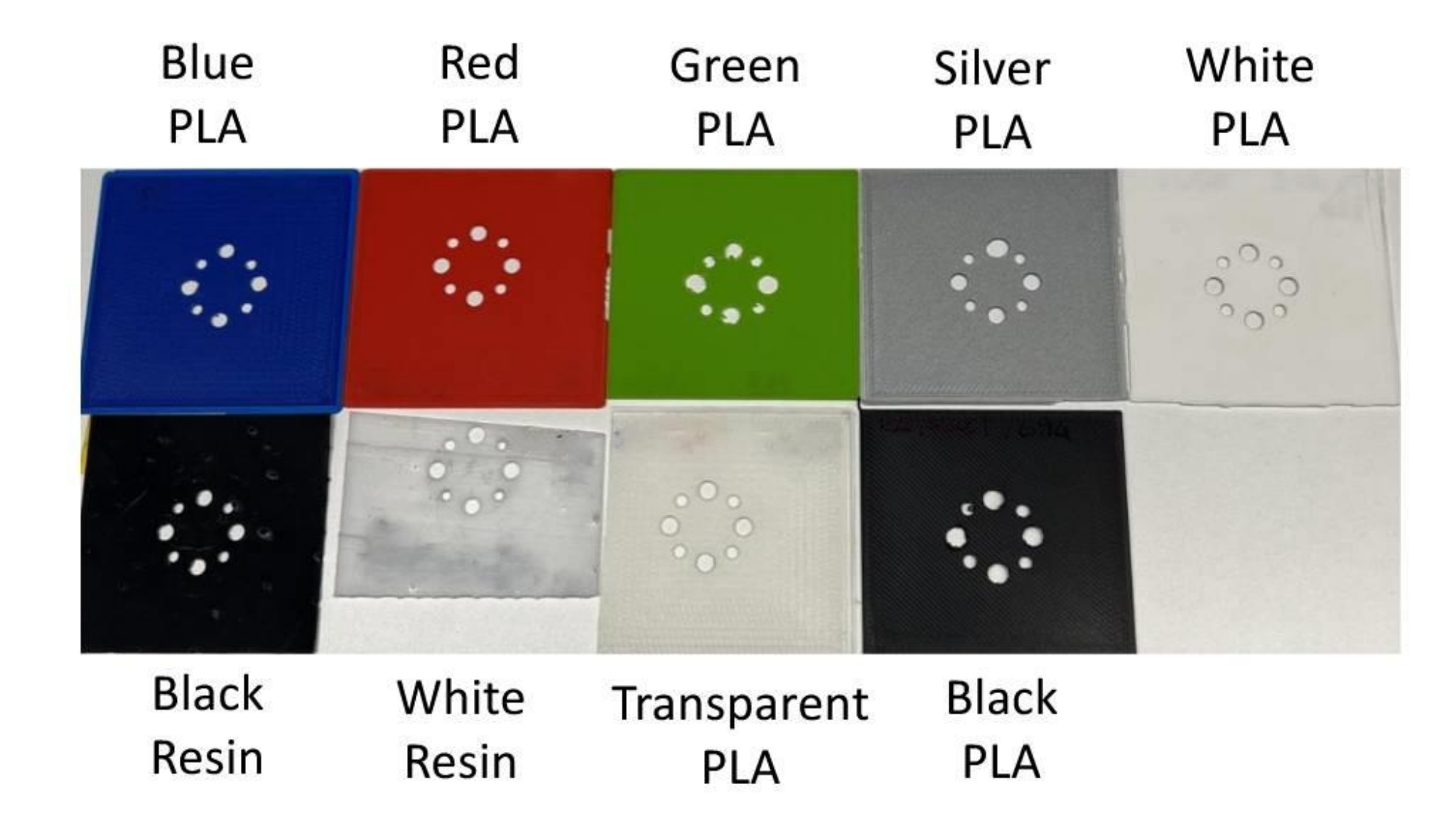}
\caption{The measured samples, including seven different types of Ultimaker
PLA and two resins produced by Formlabs. The holes in the samples do not affect the extracted permittivity values and are required due to the presence of screws tightening the waveguide flanges. }
\label{fig:sample}
\end{figure}

Figure~\ref{fig:sample} shows 3D-printed material samples comprising seven types of commercially available Ultimaker PLA, with colors identified by their commercial names (Blue, Red, Green, Silver, White, Black, and Transparent), as well as two types of resin (Black and White) produced by Formlabs. The PLA samples were fabricated using the Ultimaker S5 3D printer based on the fused deposition modeling technique (FDM). The infill pattern was set to Grid; however, since the structure is two-dimensional, the infill density parameter does not affect the result. The resins were fabricated with the Formlabs Form 2 printer based on stereolithography (SLA).

To analyze the permittivity of these materials, the S-parameters were measured using an  E8361C PNA Network Analyzer (Agilent Technologies). The VNA drives submillimeter waves extender WR10 VNAX (Virginia Diodes Inc.) to reach frequency band 70-110 GHz. We employed the Thru-Reflect-Line (TRL) calibration method to accurately remove systematic errors in our measurement setup, ensuring precise characterization of the device under test while properly defining the phase reference plane at the calibration plane. Besides, Ansys HFSS was employed to perform numerical simulations to fit the measured S-parameters. As a result of the simulations, the real and imaginary parts of the permittivity were extracted at nine frequency points between 70 GHz and 110 GHz.
During the experiment, several challenges were encountered, with the primary issue being thickness variation on the material surfaces due to non-uniform printing from the 3D printer. To mitigate thickness variations, the sample's surface region in contact with the waveguide flanges was measured at several points using a micrometer, and the measurements were subsequently averaged.  Additionally, for the PLA material samples, holes were manually drilled instead of being incorporated during printing. This approach avoids further thickness inconsistencies, as printing the holes would inevitably cause thickening near their edges.
In contrast, for the resin samples, the higher print quality ensured surface homogeneity even when the holes were included in the design (no significant impact on the measured data was observed). 

During the measurements, we found that excessive tightening of the screws can result in non-physical results (negative loss tangent). Furthermore, manual tightening should be avoided as it leads to inconsistent measurements, as non-uniform pressure could bend the sample, alter its effective thickness, and consequently affect the measured permittivity. This inconsistency posed a significant problem for the repeatability of the experiment. To resolve this issue, a torque meter was used to ensure uniform and consistent pressure of the tightened sample. Our observations revealed that material dimensions did not considerably affect the measurements unless they were smaller than the waveguide flange size. Besides, no anisotropy was observed in the printed samples, as the material properties were uniform in all directions. Isomorphism was consistent across all the materials tested. Furthermore, we observed that the number of screws used for tightening the sample between the flanges (two versus four) had a minimal influence on the results, as long as the screws were placed symmetrically and tightened with the same force.

\section{Results}
\begin{figure*}[tb]
    \centering
    \begin{minipage}{0.5\textwidth}
        \centering
        \includegraphics[width=\textwidth]{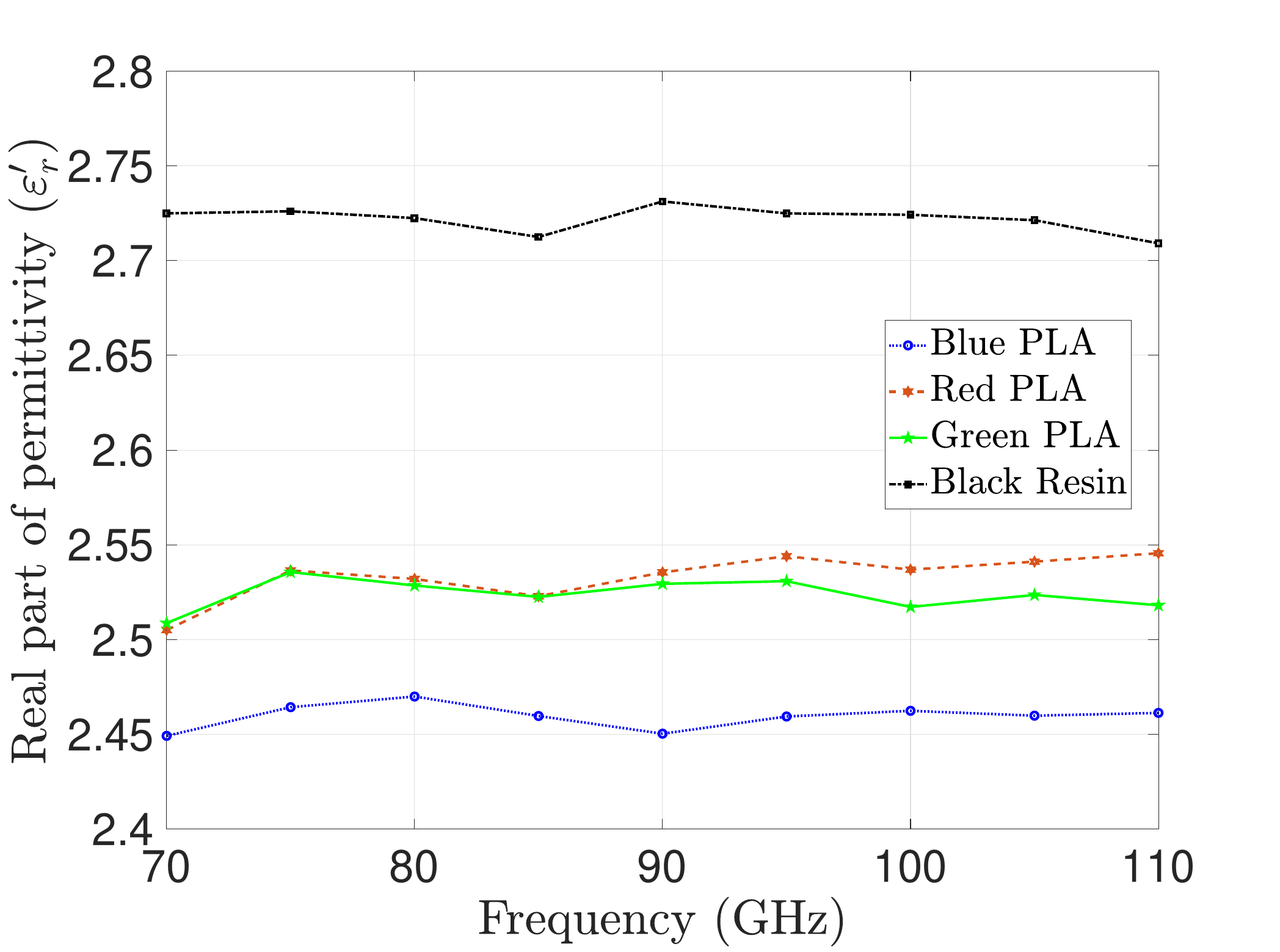}
      
    \end{minipage}%
    \begin{minipage}{0.5\textwidth}
        \centering
        \includegraphics[width=\textwidth]{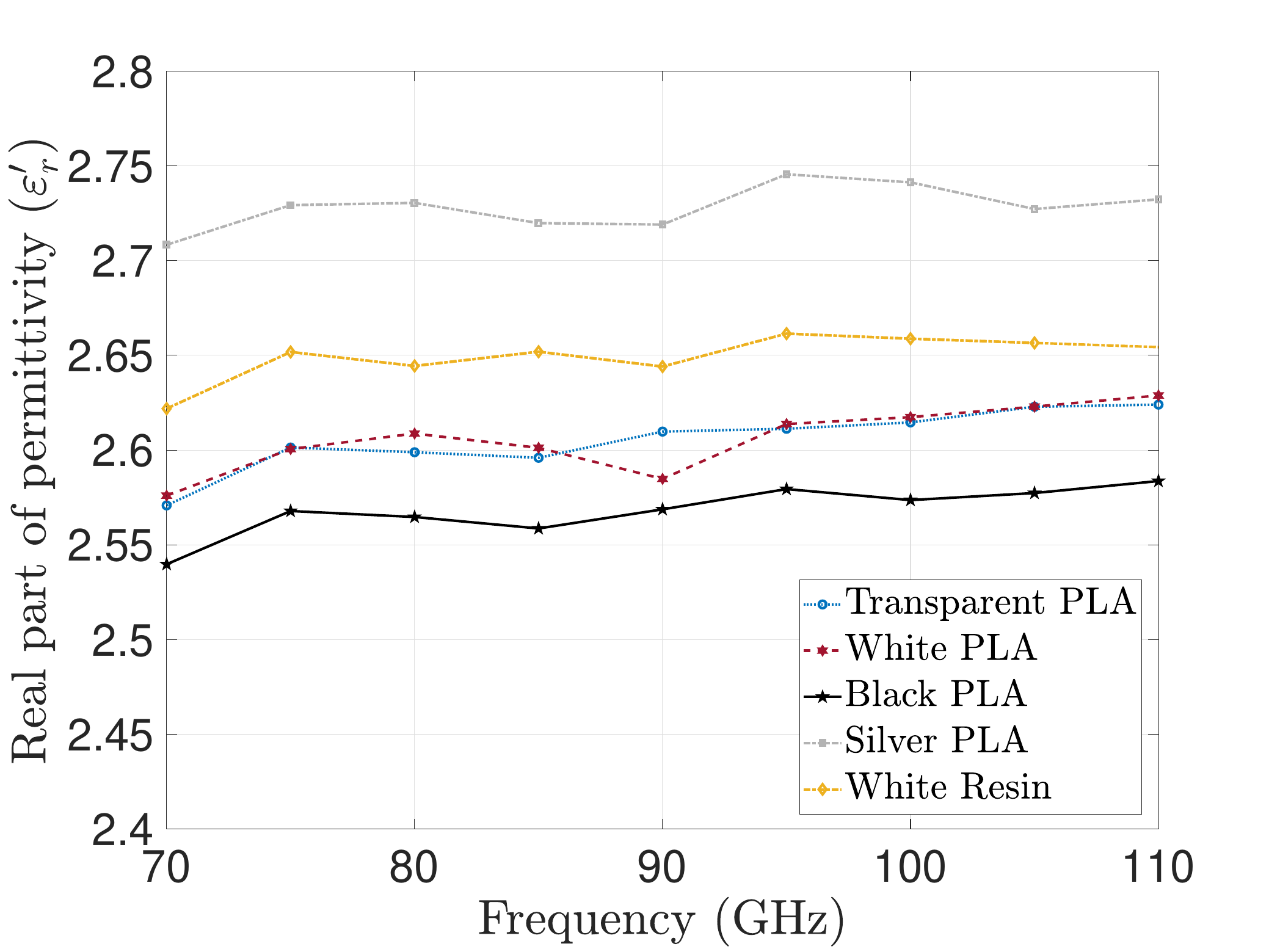}

    \end{minipage}
    \caption{Real part of permittivity versus frequency for measured samples. The plot shows relatively stable permittivity values with minor fluctuations, with Silver PLA and White Resin exhibiting higher permittivity compared to the other materials. However; the permittivity of Blue PLA is the lowest among the tested materials.}
    \label{fig:panel1}
\end{figure*}

\begin{figure*}[tb]
    \centering
    \begin{minipage}{0.5\textwidth}
        \centering
        \includegraphics[width=\textwidth]{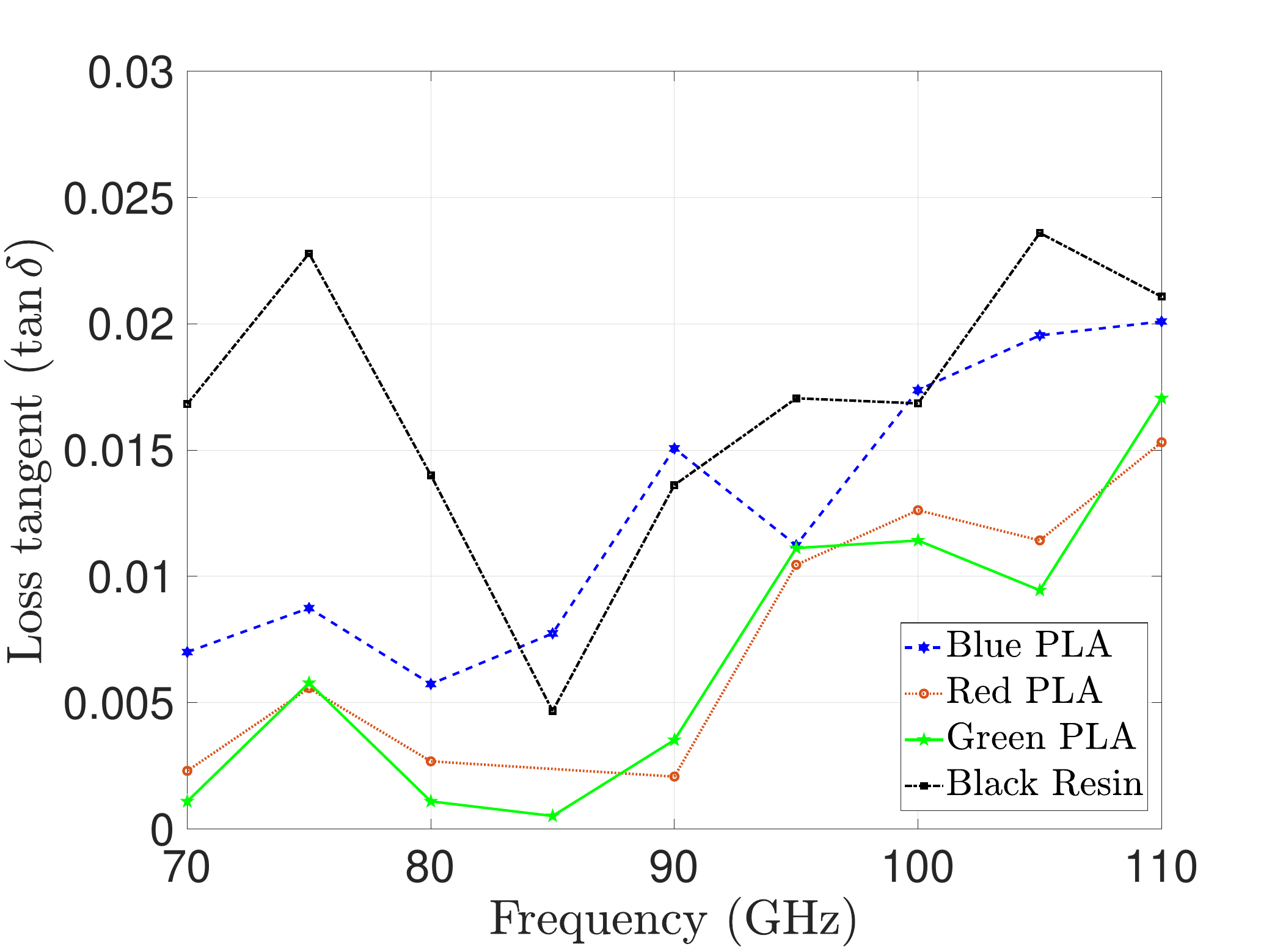}
   
    \end{minipage}%
    \begin{minipage}{0.5\textwidth}
        \centering
        \includegraphics[width=\textwidth]{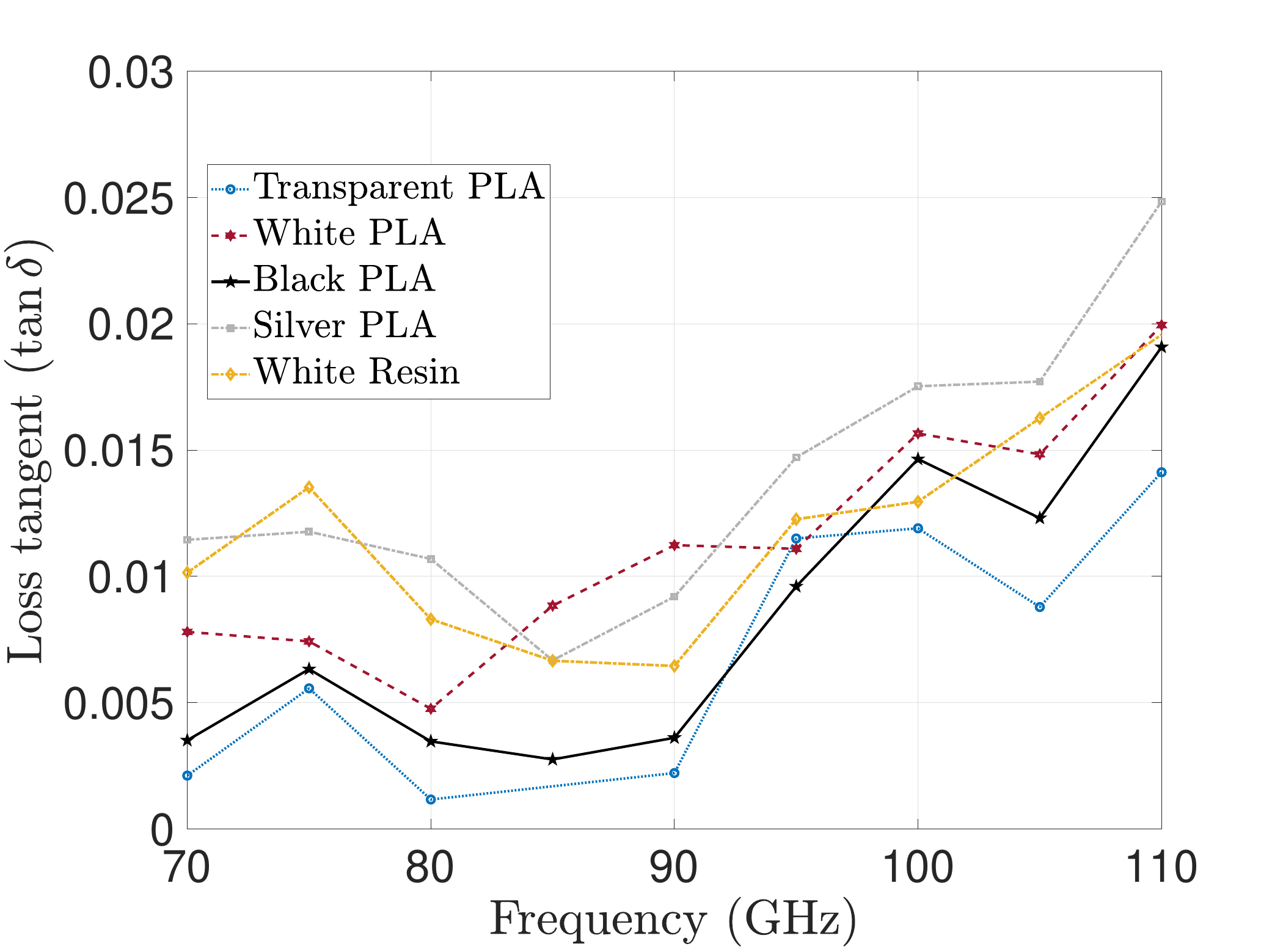}
   
    \end{minipage}
    \caption{The frequency dependency of loss tangent for measured samples.}
    \label{fig:panel2}
\end{figure*}

In this section, we present the extracted complex permittivity results between 70 GHz to 110 GHz for all the materials we measured. Figure~\ref{fig:panel1} depicts the real part of permittivity $\varepsilon_{\rm r}'= \Re (\varepsilon_{\rm r})$, while Fig.~\ref{fig:panel2} shows  the loss tangent $\tan \delta = \Im (\varepsilon_{\rm r})/ \Re (\varepsilon_{\rm r})$.  
The measured values at nine equally spaced frequency points are represented by markers. 


According to Fig.~\ref{fig:panel1},  Silver PLA and Black resin exhibit the highest permittivity values, reaching the average value above 2.7. The Blue PLA in contrast has the lowest permittivity value  near 2.45. Although minor fluctuations are observed due to the imperfections of the extraction method, the permittivity remains relatively stable across different frequencies for all materials, revealing a low normal frequency dispersion behaviour. This, in turn, implies that material resonances are far from the analyzed frequency band.

From Fig.~\ref{fig:panel2}, one can observe that Silver PLA and Black Resin exhibit the highest loss tangent values, exceeding 0.02 at higher frequencies, indicating greater dissipation loss compared to the other materials. In contrast, Transparent, Green, and Red PLA demonstrate the lowest loss tangent values, making them more suitable for applications that require minimal dissipation loss at high frequencies. Overall, the loss tangent values for all materials tend to increase with frequency, indicating the presence of material resonances at higher frequencies. For some materials, such as Black Resin, significant fluctuations are observed at certain frequency points. By looking at Figs.~\ref{fig:panel1} and \ref{fig:panel2}, one can observe that the extracted loss tangent exhibits stronger deviations from the mean value. This behavior arises because, when the wave propagates through the low-loss dielectric material slab, the attenuation is insufficiently accumulated.

Consequently, uncertainties in the S-parameters and subsequently in $Y_{\rm p}$ can lead to inaccuracies in the estimation of the loss tangent~\cite{1159663}.

\section{Conclusion}\label{sec:conclusion}

In this paper, we investigated the real part of the permittivity and the loss tangent values of several commercially available 3D-printable materials across the 70--110 GHz frequency range. These materials are widely used in both FDM and SLA additive manufacturing technologies. The results were obtained through experimental measurements of the shunt admittance, with permittivity extracted using the open-waveguide method. Our findings indicate that all analyzed materials remain non-resonant and exhibit low losses within the studied frequency range, making them promising candidates for microwave and millimeter-wave applications.

Notably, although many of the analyzed materials belong to the same brand and differ only in color, our measurements revealed significantly distinct permittivity values. Frequency-averaged real parts of permittivity ranged from as low as 2.45 to as high as 2.75 for different materials. Furthermore, the loss tangent values varied substantially across the samples of different colors.

Characterizing the frequency range of 70--110 GHz is particularly challenging due to the limitations of conventional measurement techniques. However, this range is increasingly relevant as future wireless communication technologies, including 6G, push toward higher frequencies. Such advancements necessitate the development of novel devices (e.g., metasurfaces~\cite{asgari2024multifunctional} and antennas~\cite{malik2019antenna,wang20233d}) and low-loss, cost-effective materials for their construction. The approach presented here can be readily applied to characterize the complex permittivity of many other 3D-printable materials in the sub-THz frequency range.

\section*{Declarations}

\subsection*{Ethical Approval}
We confirm that the manuscript has been read and approved by all named authors and that there are no other persons who satisfy the criteria for authorship but are not listed. We further confirm that the order of authors listed in the manuscript has been approved by all of us.
\subsection*{Conflict of Interest}
The authors declare no competing interests.
\subsection*{Funding}
This research is supported by the Research Council of Finland within the RCF-DoD Future Information Architecture for IoT initiative under grant no. 365679, the Finnish Foundation for Technology Promotion.
\subsection*{Availability of Data and Materials}
The authors declare that the data supporting the findings of this study are available within the paper and from the corresponding authors upon request.

\ifCLASSOPTIONcaptionsoff
  \newpage
\fi

\bibliographystyle{IEEEtran}

\bibliography{Bibliography}

\vspace{2cm}

\textbf{Kamil A. Işık}
received the B.Sc. degrees in Electrical and Electronics Engineering and Mathematics from Middle East Technical University, Ankara, Türkiye, in 2025 and 2026, respectively. He was a visiting research assistant at Aalto University in the summer of 2024. During that time, he studied on 3D printing materials, HFSS simulations, and millimeter-wave measurements.
\vspace{1cm}

\textbf{Mohammad M. Asgari}
received his B.Sc. degree in Electrical
Engineering (Telecommunications) from Babol Noshirvani Uni-
versity of Technology, Mazandaran, Iran in 2018 and his M.Sc. in
Electrical Engineering (Field and Waves) from the Sharif Univer-
sity of Technology, Tehran, Iran in 2021. He has been a doctoral
researcher in the Department of Electronics and Nanoengineering
at Aalto University, Espoo, Finland, since 2022. His main research
topics are metamaterials, time-varying systems, and inverse design.

\vspace{1cm}

\textbf{Xuchen Wang}
received a B.Sc. degree in optical information science and technology from Northwestern Polytechnical University,
Xi’an, China, in 2011, a master’s degree from the Department
of Optical Engineering, Zhejiang University, Hangzhou, China, in
2014, and a Ph.D. degree (Hons.) from the Department of Elec-
tronics and Nanoengineering, School of Electrical Engineering,
Aalto University, Aalto, Finland, in 2020. He worked as a Radio
Frequency Engineer at Huawei (Shanghai, China), and TP-Link
(Shenzhen, China) from 2014 to 2016. He worked as a Post-Doctoral Researcher
at Karlsruhe Institute of Technology, Germany from 2022 to 2023. He is currently
working as a Professor in the Harbin Engineering University, China.

\vspace{1cm}

\textbf{Maxim Masyukov}
Maxim Masyukov is a PhD student at the Department of Electronics and Nanoengineering at Aalto University.  He has been a visiting student researcher with Chalmers University of Technology, Gothenburg, Sweden in 2022 and NASA Jet Propulsion Laboratory, California Institute of Technology, USA. His research interests include cryogenically cooled devices and instrumentation for sub-mm-wave astronomy.

\vspace{1cm}

\textbf{Irina Nefedova}
Irina Nefedova was born in Saratov, Russia, in 1989. She received the B.Sc. and M.Sc. degrees in physics from Saratov State University, Russia, in 2010 and 2012, respectively, and the Lic.Sc. and D.Sc. degrees in electrical engineering from Aalto University, Espoo, Finland, in 2015 and 2017, respectively.,She has been a Postdoctoral Researcher and Visiting Scientist with the University of Hamburg, Hamburg, Germany (2018), the Jet Propulsion Laboratory (JPL), California Institute of Technology (Caltech), Pasadena, CA, USA (2019), and the Chalmers University of Technology, Gothenburg, Sweden (2020–2021). Since 2019, she has been working with Aalto University, first as a Postdoctoral Researcher and later as a Project Specialist. Her research interests include millimeter-wave, THz, and quasi-optical measurements with applications in remote sensing and medical

\vspace{1cm}

\textbf{Zachary Taylor}
Zachary Taylor (Member, IEEE) received the B.S. degree in electrical engineering from the University of California, Los Angeles (UCLA), Los Angeles, CA, USA, in 2004, and the M.S. and Ph.D. degrees in electrical engineering from the University of California, Santa Barbara, Santa Barbara, CA, USA, in 2006 and 2009, respectively.,From 2013 to 2018, he was an Adjunct Assistant Professor with appointments with the Department of Bioengineering, Department of Electrical Engineering, and Department of Surgery, UCLA. From 2018 to 2022, he was an Assistant Professor with the Department of Electronics and Nanoengineering, Aalto University, Espoo, Finland. Since 2022, he has been an Associate Professor with the same department. His current research interests include submillimeter-wave and THz imaging and sensing and THz frequency calibration techniques for antenna measurements, personnel imaging, and clinical diagnostics.

\vspace{1cm}

\textbf{Viktar S. Asadchy}
received his diploma and M.Sc. degrees in
Physics from Gomel State University, Belarus, in 2013 and 2014,
respectively. In 2017, he obtained his D.Sc. degree in Electrical
Engineering from Aalto University, Finland. From 2019 to 2022,
he served as a Postdoctoral Fellow at the Department of Electri-
cal Engineering, Stanford University, CA, USA. He is currently an
Assistant Professor in the Department of Electronics and Nanoengi-
neering at Aalto University, Finland. He is an Elected Associate
Member of URSI. His primary research interests include metasurfaces, reconfig-
urable intelligent surfaces, metamaterials, photonic crystals, time-varying systems, and nanophotonics.

\clearpage

\end{document}